\RequirePackage[2020-02-02]{latexrelease}
\documentclass[prl,preprintnumbers,onecolumn,aps,superscriptaddress,floatfix]{revtex4}
\usepackage{graphicx,soul}
\usepackage{graphicx,dcolumn,bm,epsfig,amsmath,amssymb,textcomp,}
\usepackage{float}
\usepackage{wrapfig}
\usepackage[utf8x]{inputenc}

\usepackage{xcolor}
\usepackage{bbm}

\begin{document}

\title{Evolution of energy density fluctuations in the presence of a magnetic field}

\author{Shreyansh S. Dave\footnote{ssdave90@gmail.com}}
\affiliation{Department of Physics, J.L.N. College, Dehri-On-Sone (Rohtas) 821307, Veer Kunwar Singh University, Ara, India}
\author{Subrata Pal\footnote{spal@tifr.res.in}}
\affiliation{Department of Nuclear and Atomic Physics, 
Tata Institute of Fundamental Research, Mumbai 400005, India}

\begin{abstract}

In this proceeding, we study the evolution of energy density fluctuations in the presence of a static and uniform magnetic field. By numerically solving the relativistic Boltzmann-Vlasov equation within the relaxation time approximation and performing the momentum mode analysis of different wavelength fluctuations, we show that the magnetic field increases the damping of mode oscillations. This causes a qualitative change in the fluctuations present in the system at the timescale required to achieve a local equilibrium state.
\end{abstract}

\maketitle

\noindent 

\section*{I. Introduction}

In non-central relativistic heavy-ion collisions a strong magnetic field can be generated due to the colliding nuclei \cite{larry, ourMHD}. Such magnetic field can survive for a longer time in highly electric conducting medium of quark, antiquark, and gluons formed in the participant zone \cite{ranjita, tuchin0, KharzMHD, ourMHD, ourRev, mhd1, mhd2}. In the initial stages of the collision, the partonic medium becomes in the out-of-equilibrium state, having large gradient fluctuations over the participant zone \cite{hydroSim, Schenke12}. As this partonic medium approaches the local equilibrium state, the fast-evolving short-wavelength fluctuations die out quickly and the energy density of the medium forms a smooth structure. However, relatively long-wavelength fluctuations survive in energy density which further participate in the hydrodynamic evolution of the system and generate various higher order flow-harmonics of the detected hadrons \cite{kapusta1, spal1, stephanov1}. During the process of thermalization, the strong magnetic field can also affect the dynamics of various wavelength-fluctuations and result in distinct initial conditions for dynamical evolution of the quark-gluon plasma as compared to the medium dynamics without any magnetic field.

In this work, we study the effects of a static and uniform magnetic field ${\bf B}$ on the evolution of energy density fluctuations of different wavelengths by numerically solving  the Boltzmann-Vlasov equation in the relaxation time approximation (RTA) \cite{rischke1, rischke2, Ashu1, bgk, Anderson, PaulR}. With a detailed momentum mode analysis of fluctuations, we explicit show that as the system approaches a local equilibrium state, the fluctuations dynamics is affected by the magnetic field which increases the damping of mode oscillations in the transverse direction to ${\bf B}$ \cite{bveq}. In particular, the magnetic field induces a much smoother energy density profile and larger fluctuation strength as compared to field-free case.

\section*{II. Boltzmann-Vlasov Equation}

In the relaxation time approximation (RTA) \cite{bgk, Anderson, PaulR, Ashu1}, the system is considered to be near the equilibrium state. Hence, the single-particle phase-space distribution function $f\equiv f(t,{\bf x},{\bf k})$ of particles can be written as $f=f_{\rm eq} + \delta f$, where $f_{\rm eq}$ is the local equilibrium distribution function of the system and $\delta f \ll f_{\rm eq}$ is the nonequilibrium deviation. In this approximation, evolution of the distribution function is given by \cite{rischke2}
\begin{equation}
 k^\mu \partial_\mu f + q F^{\mu \nu} k_\nu
 \frac{\partial}{\partial k^\mu} f = \frac{k^\mu u_\mu}{\tau_c}
(f_{\rm eq}-f).
\label{BV:eq}
\end{equation}
Here $k^\mu = (E_k,{\bf k})$ is the four-momentum of particle, $F^{\mu \nu}$ the electromagnetic field tensor, $q$ the electric charge of particle, $\tau_c$ the relaxation time for local equilibration, $u^\mu = \gamma (1,{\bf v})$ the four-velocity of the fluid, and $\gamma = 1/\sqrt{1-v^2}$ the Lorentz factor.

We consider a static and uniform magnetic field along $y$ direction given by ${\bf B} = B_0 \hat y$. Since this magnetic field does not have any effect along $y$ and $k_y$ directions of the distribution function, we consider the evolution of a (3+1)-dimensional distribution function $f \equiv f(t, x, k_x , k_z)$ with $k_y = 0$, where inhomogeneity along $y$ and $z$ directions is ignored. The Boltzmann-Vlasov (BV) equations for particles and anti-particles in the dimensionless form \cite{bveq} can be written as
\begin{equation}
\begin{split}
 E_k'\frac{\partial f}{\partial t'} +
 k'_x \frac{\partial f}{\partial x'} +
 \beta_0 \Big(k_x' \frac{\partial f}{\partial k_z'} -
 k_z' \frac{\partial f}{\partial k_x'} \Big) =
 k'^\mu u_\mu (f_{\rm eq} - f),
 \\
 E_k'\frac{\partial \bar{f}}{\partial t'} +
 k'_x \frac{\partial \bar{f}}{\partial x'} -
 \beta_0 \Big(k_x' \frac{\partial \bar{f}}{\partial k_z'} -
 k_z' \frac{\partial \bar{f}}{\partial k_x'} \Big) =
 k'^\mu u_\mu (\bar{f}_{\rm eq} - \bar{f}),
 \end{split}
 \label{eomF}
\end{equation}
where the dimensionless variables are defined as $t' = t/\tau_c$, $x' = x/\tau_c$, ${\bf k'} = {\bf k}/m_0$ (hence $E_k' = E_k /m_0$), with $|{\bf k}| =\sqrt{k_x^2+k_z^2}$ and (anti)particle's mass $m_0$. We vary the \textquotedblleft magnetic field parameter" $\beta_0 = |qB| \tau_c/m_0$ to study the impact of magnetic field $B$ on the evolution of fluctuations. Medium velocity can be determined by Landau frame definition that also satisfies Landau matching condition required for energy-momentum and net-particle current conservations \cite{Anderson, sunil2022, spal2, chandro22}. We generate an initial nonequilibrium configuration of $f(x', k_x', k_z')$ and $\bar{f}(x', k_x', k_z')$ and solve numerically the two coupled differential equations \eqref{eomF}. We also vary the wavelength of fluctuations to study the impact of magnetic field on various fluctuation modes of energy density; the simulation details can be found in Ref. \cite{bveq}.

\section*{III. Simulation Results}

\begin{wrapfigure}{r}{0.3\textwidth}
  \begin{center}
    \includegraphics[width=0.3\textwidth]{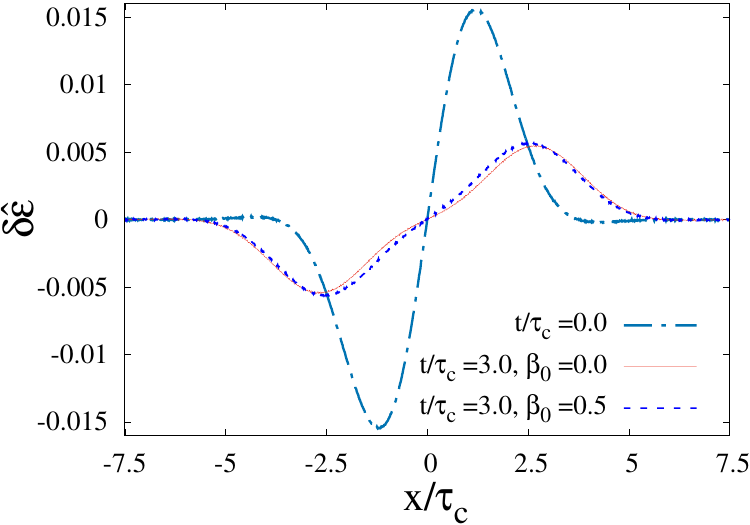}
  \end{center}
  \caption{Energy density fluctuations $\delta \hat\varepsilon$ at time $t/\tau_c=0$ and $t/\tau_c=3.0$ in the absence and presence of ${\bf B}$. Figure is from \cite{bveq}.}
  \label{fig1}
\end{wrapfigure}

By solving the BV equations, we obtain the distribution functions of particles and anti-particles which are used to calculate the \textquotedblleft dimensionless" energy density of the system. The energy density fluctuations can be extracted by using the relation
\begin{equation}
\delta \hat\varepsilon = \frac{1}{\hat\varepsilon_0} \{
\hat\varepsilon(x',t') - \langle \hat\varepsilon(x',t')\rangle \},
\end{equation}
where $\hat\varepsilon_0$ is the global equilibrium energy density and $\langle \hat\varepsilon(x',t') \rangle$ is the spatial average of $\hat\varepsilon (x',t')$. Figure \ref{fig1} shows the energy density fluctuations at time $t/\tau_c=0$ and $t/\tau_c=3.0$ in the absence and presence of the magnetic field. It clearly indicates that the magnetic field affects the fluctuations in the charged medium evolving transverse to ${\bf B}$.

\begin{wrapfigure}{r}{0.3\textwidth}
  \begin{center}
    \includegraphics[width=0.3\textwidth]{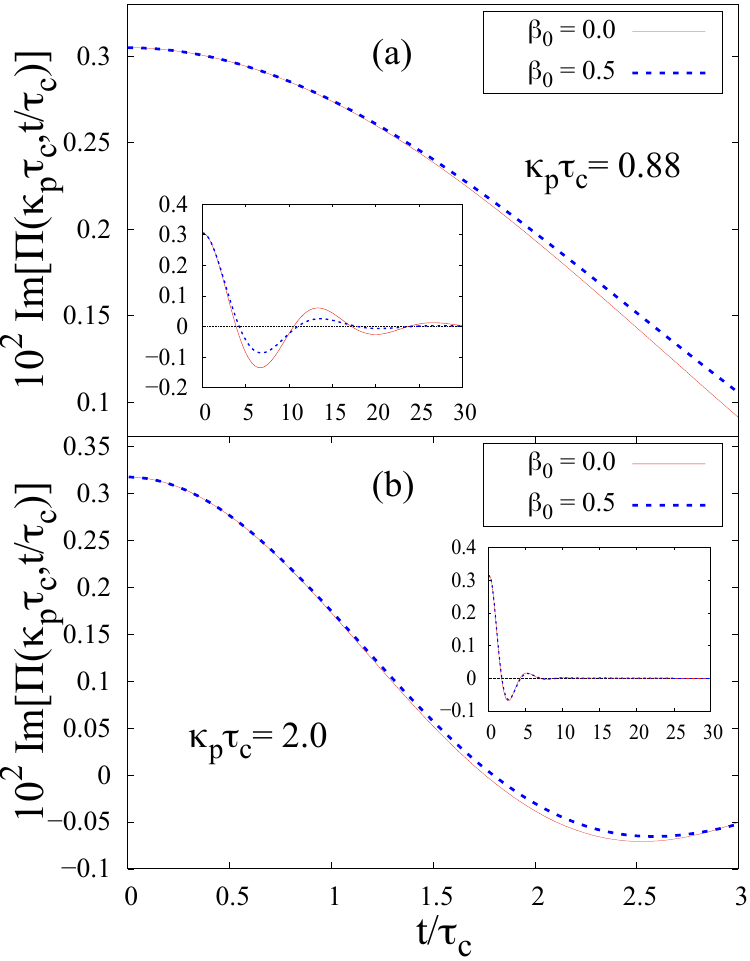}
  \end{center}
  \caption{Time evolution of the most dominant mode of energy density fluctuations (a) for $\kappa_p \tau_c=0.88$ and (b) for $\kappa_p \tau_c=2.0$ in the absence and presence of magnetic field. Evolution of these modes for a longer time are shown in insets. Figure is from \cite{bveq}.}
  \label{fig2}
\end{wrapfigure}

\begin{figure}[b]
\includegraphics[width=0.3\linewidth]{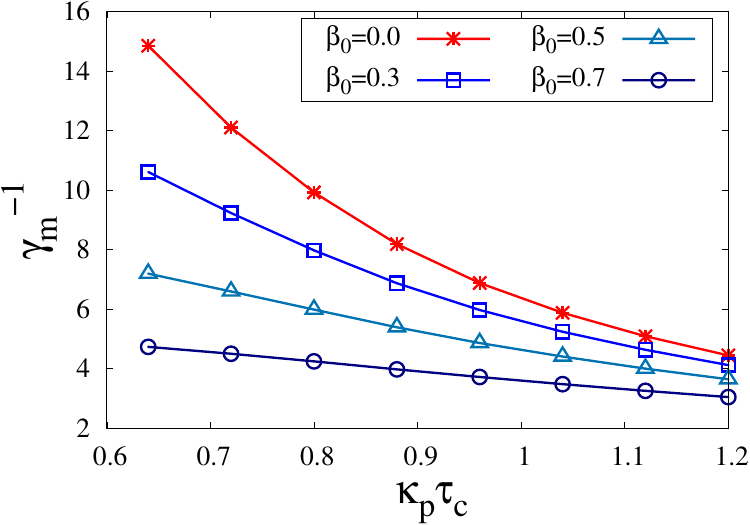}~~~
\includegraphics[width=0.252\linewidth]{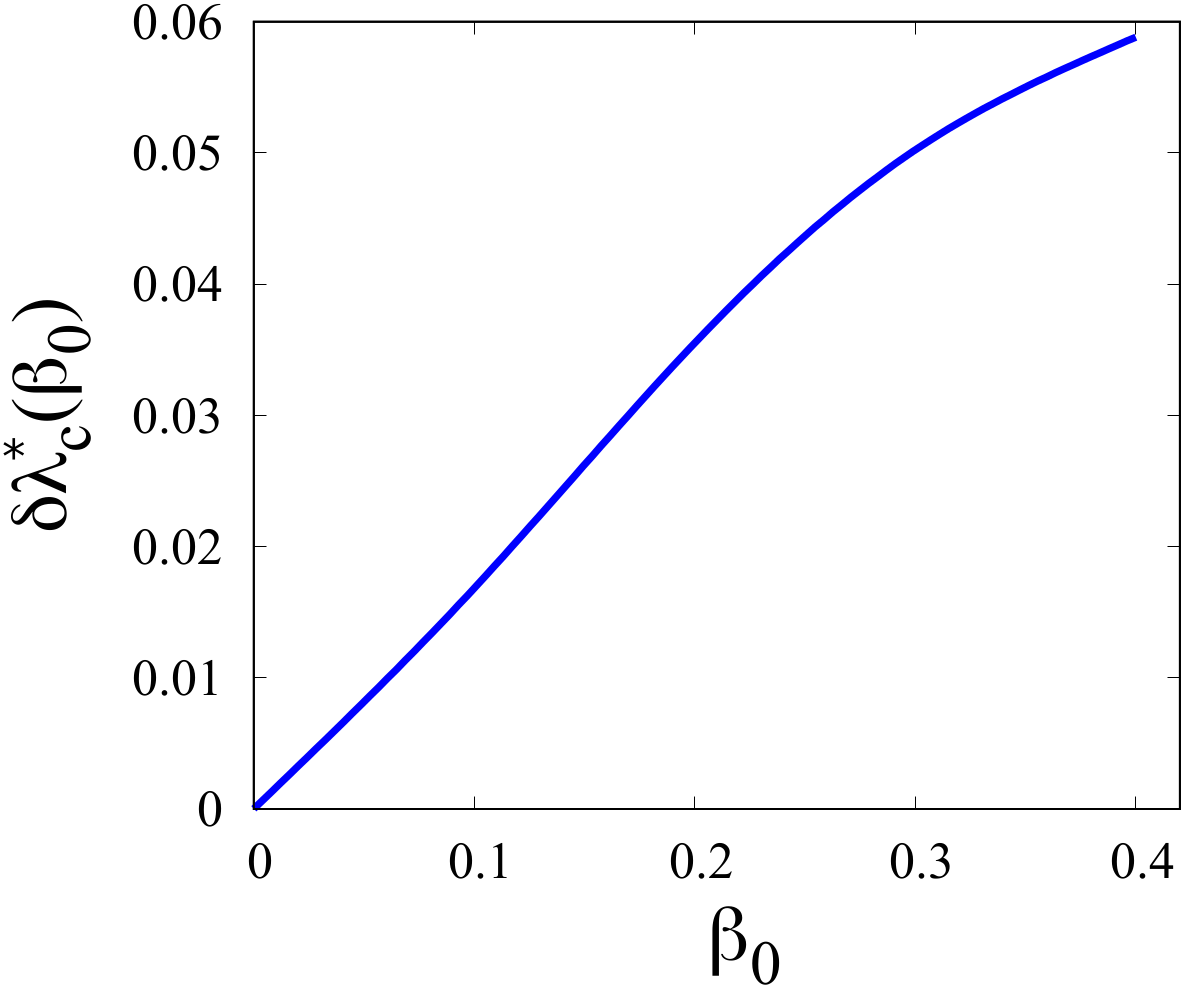}~~~
\caption{Left: The decay timescale $\gamma_m^{-1}$ of a mode as a function of mode-momentum $\kappa_p \tau_c$ for different values of $\beta_0$. Right: The fractional change of wavelength cutoff $\delta \lambda^*_c$ as a function of $\beta_0$. Figures are from \cite{bveq}.}
 \label{fig3}
\end{figure}

To analyse the effects of magnetic field, the energy density fluctuations are Fourier transformed from configuration $x'$ space to the momentum $\kappa'$ space as
\begin{equation}
  \Pi(\kappa',t') = \frac{1}{L}\int_{-L/2}^{L/2} dx'~
  \delta \hat\varepsilon(x',t')~e^{i \kappa' x'},
  \label{eq:four}
\end{equation}
where $\kappa' = \kappa \tau_c$ is the dimensionless momentum of the mode $\Pi(\kappa',t')$, $\kappa$ the dimensionful momentum, and $L$ the system size. We focus on the time evolution of the most dominant mode of energy density fluctuations whose momentum is denoted as $\kappa'_p = \kappa_p \tau_c$. Figure \ref{fig2} displays the time evolution of Im$[\Pi (\kappa'_p, t')]$ for (a) $\kappa_p \tau_c=0.88$ and (b) $\kappa_p \tau_c=2.0$ in the absence and presence of the magnetic field. Figure \ref{fig2}(a) corresponds to the fluctuations shown in Fig. \ref{fig1}, while Fig. \ref{fig2}(b) corresponds to a relatively short wavelength fluctuation. It is clear from these plots that the high momentum (short wavelength) mode is quickly suppressed by the magnetic field as compared to a low momentum mode. It is also evident from Fig. \ref{fig2} that the time evolution of ${\rm Im}[\Pi(\kappa'_p,t')]$ can be expressed by the damped harmonic oscillator function:
\begin{equation}
 \text{Im}[\Pi(\kappa'_p,t')] = \alpha_0 \cos(\omega t' - \phi)
 \exp(-\gamma_m t').
\label{eq:dhf}
\end{equation}
Here $\alpha_0 \equiv \alpha_0(\kappa'_p)$ is the amplitude of the oscillator, $\omega \equiv \omega(\kappa'_p)$ the dimensionless angular frequency, $\phi \equiv \phi(\kappa'_p)$ the phase, and $\gamma_m \equiv \gamma_m(\kappa'_p)$ the dimensionless damping coefficient. In Fig. \ref{fig3} [left panel], we show the \textquotedblleft dimensionless decay timescale" $\gamma_m^{-1}$ of dominant mode as a function of the mode-momentum $\kappa_p\tau_c$ for different values of magnetic field parameter $\beta_0$. The magnetic field decreases the decay timescale of the mode or equivalently increases the damping coefficient of mode oscillations.

Each curve in Fig. \ref{fig3} [left panel] is fitted with a power law scaling $\gamma_m^{-1} = s_0 (\kappa_p \tau_c)^{-s_1} + s_2$. We compare the decay timescale $\gamma_m^{-1} \tau_c$ of the mode with the local-equilibration timescale $\tau_c$ to obtain the ultraviolet momentum cutoff $\kappa_c \tau_c$ of fluctuations above which all the higher momentum modes are suppressed. This can be obtained by putting $\gamma_m^{-1}=1$ in the above power law scaling, which also provides the \textquotedblleft dimensionless" wavelength cutoff $\lambda^*_c=\lambda_c/\tau_c$ below which any inhomogeneity in the energy density is suppressed. In Fig. \ref{fig3} [right panel], the fractional change of wavelength cutoff $\delta \lambda^*_c(\beta_0) = [\lambda^*_c(\beta_0)- \lambda^*_c(0)] /\lambda^*_c(0)$ is plotted as a function of $\beta_0$. The wavelength cutoff $\lambda_c$ is equivalent to the coarse-grained length scale used for the hydrodynamic description for medium evolution. The increasing value of $\delta \lambda^*_c$ as a function of $\beta_0$ suggests that the magnetic field suppresses the short wavelength fluctuations present in the system causing a smoother coarse-grained structure of the hydrodynamic variables.

The above effects of magnetic field on the evolution of fluctuations arise mainly due to the Lorentz force exerted by the magnetic field along $z'$ direction on the fluctuations evolving along $x'$. Consequently, $\hat T^{0z}$ and $\hat T^{xz}$ components of energy-momentum tensor $\hat T^{\mu \nu}$ are generated leading to the suppression of momentum density $\hat T^{0x}$ along the $x'$ direction \cite{bveq}. This causes an increase in the damping coefficient of the mode oscillations as shown previously.

\begin{figure}[t]
\includegraphics[width=0.293\linewidth]{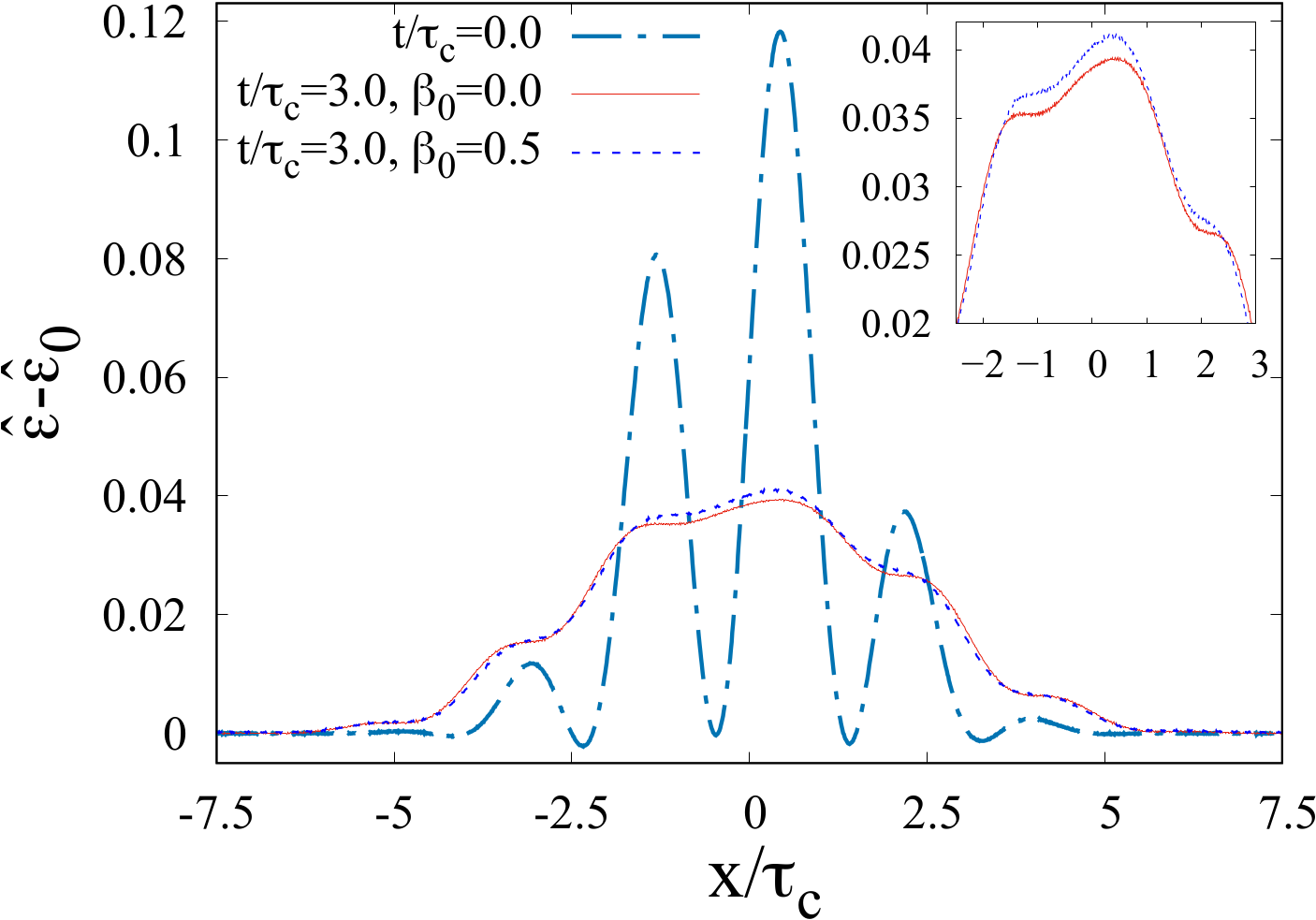}~~~
\includegraphics[width=0.3\linewidth]{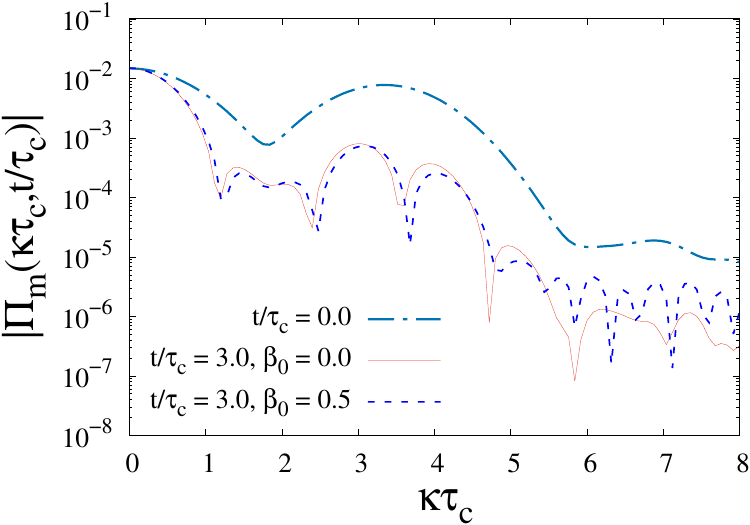}
 \caption{Left: The energy density fluctuations, $(\hat\varepsilon -  \hat\varepsilon_0)$, at time $t/\tau_c=0$ and $t/\tau_c=3.0$ in the absence and presence of the magnetic field. In the presence of the magnetic field, the energy density profile is smoother and has a larger peak value as compared to $B=0$. Right: Momentum spectrum of energy density fluctuations shown in the left figure. Figures are from \cite{bveq}.}
 \label{fig4}
\end{figure}

We next present the results for the fluctuations having a mixture of low and a very high momentum modes. In Fig. \ref{fig4} [left panel], we show the energy density fluctuations, $(\hat\varepsilon-\hat\varepsilon_0)$ for such cases at times $t/\tau_c = 0$ and $t/\tau_c = 3.0$ in the absence and presence of the magnetic field. In the presence of the magnetic field, at later time, the energy density profile becomes slightly smoother and has a larger peak value as compared to field-free case; see inset of this figure. The enhanced suppression of the short wavelength (high momentum) modes and a slow dissipation of long wavelength (low momentum) fluctuations lead to such a smoother and larger peak-value energy density profile.
In Fig. \ref{fig4} [right panel], we present the corresponding momentum spectrum of the energy density fluctuations. This clearly depicts the suppression of the high momentum dominant modes (at $\kappa \tau_c \approx $ 3.0 and 4.0) and a slow dissipation of the low momentum modes ($\kappa \tau_c < 1.2$) in the presence of the magnetic field and explains the structure seen for the energy density profile.

\section*{IV. Summary and Conclusions}

In this work we have shown that magnetic fields can induce moderately large and nontrivial effects on the evolution of energy density fluctuations during the approach of the system towards the local equilibrium state. This occurs due to Lorentz force exerted by the magnetic field in the transverse plane leading to increase the damping of fluctuation modes. This causes a reduction in the ultraviolet cutoff of fluctuations, and also leads to a larger mode-value for relatively low momentum fluctuation modes.

The study has important phenomenological implications in the context of relativistic heavy-ion collisions. In the pre-equilibrium stage of relativistic heavy-ion collisions, various short wavelength fluctuation modes are present \cite{Schenke12}. The magnetic field can increase the damping of such modes, as demonstrated in this work, and consequently modify the characteristics of short wavelength fluctuations in the reaction plane, which can further affect the collective flow harmonics $v_n(p_T) = \langle \cos n(\phi - \Psi_n)\rangle$ of the measured hadrons \cite{Bhalerao:2014xra, Giacalone:2017uqx, saumia1, Schenke12i, saumia2, ourRev,ehr}.

\section{acknowledgments}
The simulations were performed at the High Performance Computing cluster at TIFR, Mumbai. The authors acknowledge financial support by the Department of Atomic Energy (Government of India) under Project Identification No. RTI 4002.

\end{document}